\documentclass[prd,a4paper,showpacs,preprint,byrevtex]{revtex4}
\usepackage{graphicx}
\usepackage{dcolumn}
\usepackage{amsmath}
\usepackage{array}
\usepackage{bm}
\usepackage{textcomp}

\begin{document}
%\preprint{Preprint Universit\'{e} de Mons-Hainaut}

\title{On two- and three-body descriptions of hybrid mesons}

\author{Fabien \surname{Buisseret}}
\thanks{FNRS Research Fellow}
\email[E-mail: ]{fabien.buisseret@umh.ac.be}
\author{Claude \surname{Semay}}
\thanks{FNRS Research Associate}
\email[E-mail: ]{claude.semay@umh.ac.be}
\affiliation{Groupe de Physique Nucl\'{e}aire Th\'{e}orique,
Universit\'{e} de Mons-Hainaut,
Acad\'{e}mie universitaire Wallonie-Bruxelles,
Place du Parc 20, BE-7000 Mons, Belgium}

\date{\today}

\begin{abstract}
Hybrid mesons are exotic mesons in which the color field is not in its
ground state. Their understanding deserves interest from a theoretical
point of view, because it is intimately related to nonperturbative
aspects of QCD. In this work, we analyze and compare two different
descriptions of hybrid mesons, namely a two-body $q\bar q$ system with
an excited string, or a three-body $q\bar q g$ system. In particular, we
show that the constituent gluon approach is equivalent to an effective
excited string in the heavy hybrid sector. Instead of a numerical
resolution, we use the auxiliary field technique. It allows to find
simplified analytical mass spectra and wave functions, and still leads
to reliable qualitative predictions. We also investigate the light
hybrid sector, and found a mass for the lightest hybrid meson which is
in satisfactory agreement with lattice QCD and some effective models.
\end{abstract}

% 12.39.Mk    Glueball and nonstandard multi-quark/gluon states
% 12.39.Ki    Relativistic quark model
% 12.39.Pn    Potential models

\maketitle

\section{Introduction}

The study of hybrid mesons deserves much interest in theoretical as well
as in experimental particle physics. From a theoretical point of view,
these particles are interpreted as mesons in which the color field is in
an excited state. Numerous lattice QCD calculations have been devoted to
the study of hybrid mesons, in particular to the energy levels of the
gluonic field \cite{Juge,Juge2} and to the properties of the $1^{-+}$
state, which is the lightest hybrid with exotic quantum numbers (see
Refs.~\cite{Nel02,Liu06} for useful references). On the experimental
side, we can mention the recently observed $\pi_1(1600)$ \cite{Ada98},
$\pi_1(2000)$ \cite{Lu} and $Y(4260)$ \cite{aub05}, which could be
either hybrid mesons, or tetraquark states \cite{Klem04}.

Apart from lattice QCD, hybrid mesons have been studied with
effective models for a long time. For example, we can quote the flux
tube model \cite{flux}, models with constituent gluons
\cite{horn,constg}, or the MIT bag model \cite{bag}. In potential
models, to which our paper is devoted, there are two main approaches. In
the first one, the quark and the antiquark are linked by a string, or
flux tube, which simulates the exchange of gluons responsible for the
confinement. If the string is in the ground state, it reduces to the
usual linear confinement potential for heavy quarks, and to a more
general flux tube model for light quarks, where the dynamics of the
string cannot be neglected \cite{laco89}. In this stringy picture, it is
possible for the flux tube to fluctuate at the quantum level, and thus
to be in an excited state. These string excitations are analog to the
gluon field excitations in full QCD. They have been studied for example
in Refs.~\cite{Allen:1998wp,luscher}. In the second approach, it is
assumed
that the hybrid meson is a three-body system, formed of a quark, an
antiquark, and a constituent gluon, which represents the gluonic
excitation. Two fundamental strings then link the gluon to the quark and
to the antiquark. This picture has been studied in Ref.~\cite{constg},
but also in more recent works \cite{Kalashnikova:2002tg,hyb1}.

Nowadays, the spinless Salpeter Hamiltonian (SSH) with a linear
confinement is a widely used and successful framework to compute hadron
spectra in potential models (see previous references). Since its kinetic
operator is a semi-relativistic one, most of the results are numerically
obtained. However, the auxiliary field (AF) technique, also known as the
einbein field method, allows to greatly simplify the calculations
\cite{Sem03,Simo1}. In a previous work \cite{hyb1}, it is shown that the
AF method leads to analytic solutions for the eigenvalues and wave
functions of the two-body SSH. Even if they are approximations, these
solutions are qualitatively in agreement with well-known experimental
facts, the Regge trajectories for example. The AF technique was also
applied to the case of hybrid mesons with a constituent gluon and two
static quarks. This model is able to reproduce some lattice results
concerning the gluonic energy levels and the heavy hybrid spectroscopy.
Moreover, it suggests a correspondence between the excited
flux tube and the constituent gluon approaches \cite{hyb1}. The purpose
of the present paper is now to apply the AF technique to the full
$q\bar q g$ system, without demanding that the quark and the antiquark
are static, in order to obtain formula which are valid for arbitrary
quark mass. Moreover, we will further investigate the links between the
excited string and the constituent gluon pictures. Eventually, we will
show that these approaches are equivalent for heavy quarks. Our
formalism will also allow us to study whether this equivalence is
modified in the light quark sector or not.

Our paper is organized as follows. In Sec.~\ref{afintro}, we review
the main properties of the AF method by applying it to the simple
case of mesons. In Sec.~\ref{hybexci}, we discuss the excited flux
tube model of hybrid mesons. Then, we analytically solve the $q\bar q g$
three-body
problem corresponding to hybrid mesons with a constituent gluon in
Sec.~\ref{hybcons}, and give some physical results concerning their mass
and structure. In Sec.~\ref{eff2pot}, we derive the effective
quark-antiquark potential for the $q\bar q$ system within a hybrid
meson, and we discuss its
links to the excited string picture. Finally, we compare our results to
lattice QCD in Sec.~\ref{discuss}, and draw some conclusions in
Sec.~\ref{conclu}.

\section{Mesons and auxiliary fields}\label{afintro}

A system made of two hadrons interacting through a linear confinement
can be described by the following SSH
\begin{equation}\label{ham1}
H=\sqrt{\bm{p}^{\, 2}+m^2_1}+\sqrt{\bm{p}^{\, 2}+m^2_2}+ar,
\end{equation}
where $\bm{p}^{\, 2}_1=\bm{p}^{\, 2}_2=\bm{p}^{\, 2}$ since we work in
the center of mass frame. The linear confinement can be understood as
the static contribution of a straight string, or flux tube, of tension
$a$, linking the quark and the antiquark \cite{laco89}.
In order to get rid of the square roots appearing in
Hamiltonian~(\ref{ham1}), let us now introduce three AF: Two for the
quarks, denoted
$\mu_i$, and one for the potential, $\nu$. Hamiltonian~(\ref{ham1}) then
becomes
\begin{equation}\label{ham2}
H(\mu_i,\nu)=\frac{\bm{p}^{\, 2}+m^2_1}{2\mu_1}+\frac{\mu_1}{2}+  \frac{
\bm{p}^{\, 2}+m^2_2}{2\mu_2}+\frac{\mu_2}{2}+\frac{a^2 r^2}{2\nu}+\frac{
\nu}{2}.
\end{equation}
Although being formally simpler, $H(\mu_i,\nu)$ is equivalent to $H$ up
to the elimination of the AF thanks to the constraints
\begin{subequations}\label{elim}
\begin{eqnarray}
  \delta_{\mu_i}H(\mu_i,\nu)&=&0\ \Rightarrow\ \mu_{i0}=\sqrt{\bm{p}^{\,
  2}+m^2_i},\\
  \delta_{\nu}H(\mu_i,\nu)&= &0\ \Rightarrow\ \nu_{0}=ar.
\end{eqnarray}
\end{subequations}
It is worth mentioning that $\left\langle \mu_{i0}\right\rangle$ can be
seen as a dynamical mass of the quark whose current mass is $m_i$, while
$\left\langle \nu\right\rangle$ is in this case the static string energy
\cite{Fab1}. Relations~(\ref{elim}) show that the AF are, strictly
speaking, operators. However, the calculations are considerably
simplified if the AF are considered as real numbers, and finally
eliminated by a minimization of the masses \cite{Sem03}. The extremal
values of $\mu_i$ and $\nu$, considered
as numbers,
are logically close to the values $\left\langle \mu_{i0}\right\rangle$
and $\left\langle \nu_0\right\rangle$ given by relations~(\ref{elim}).
This procedure leads to a spectrum which is an upper bound of the ``true
spectrum" (computed without AF), the differences being about $10\%$
\cite{Fab3}.

Using the AF, Hamiltonian~(\ref{ham1}) turns out to be formally a
simple nonrelativistic harmonic oscillator~(\ref{ham2}). Its mass
spectrum and wave functions read
\begin{equation}\label{mass2af}
M(\mu_i,\nu)=\omega(2n+\ell+3/2)+\frac{m^2_1}{2\mu_1}+\frac{m^2_2}{2
\mu_2}+\frac{\mu_1+\mu_2+\nu}{2},
\end{equation}
\begin{equation}\label{phico}
\psi=\phi_{n,\ell}(\beta^{1/2} r)Y^{m}_{\ell}(\theta,\varphi),
\end{equation}
where
\begin{equation}
\omega= \sqrt{a^2/\tilde{\mu}\nu},\quad \beta=\sqrt{\tilde{\mu}
a^2/\nu},\quad \tilde\mu=\frac{\mu_1\mu_2}{\mu_1+\mu_2},
\end{equation}
and where $\phi_{n,\ell}(\beta^{1/2} r)$ is a normalized radial
eigenfunction of the three dimensional harmonic oscillator.

In the special case of a light meson, one can set
$m_1=m_2=0$. Then $\mu_1=\mu_2=\mu$, and one obtains after the
elimination of the AF \cite{hyb1}
\begin{equation}\label{mass2_4}
  M^2_{ll}=8a(2n+\ell+3/2).
\end{equation}
At large angular momentum, it appears that the square mass increases
linearly with $\ell$. Thus, our solution qualitatively reproduces the
Regge trajectories, which are the best known experimental fact
concerning the light meson spectroscopy. The Regge slope is here given
by $8a$ instead of $2\pi a$, which is the exact value of the Regge slope
in the flux tube model \cite{laco89}. This is related to the AF
technique itself which gives good qualitative results, but overestimates
the masses. More precisely, the error on the exact value increases with
the number of AF introduced, as shown in Ref.~\cite{hyb1}. As a
check of this point, we can mention that with only one AF, the Regge
slope can be computed to be $7a$ \cite{qse}. What can be done to cure
this artifact of the AF method is to rescale $a$: as we know that the
exact Regge slope is around $2\pi\sigma$, with the standard value
$\sigma\approx0.2$ GeV$^2$, let us set $a=2\pi\sigma/8$. Then, a mass
formula such as expression~(\ref{mass2_4}) is able to correctly
reproduce the experimental Regge slope of the mesons \cite{laco89}.
In the following, a rescaling of the string tension will be used
to improve the results given by the AF calculations.

An other interesting case is a system composed of a light quark and of a
heavy quark. One finds that, for such a system \cite{hyb1},
\begin{equation}\label{masshl0}
  (M_{hl}-m_2)^2=4a(2n+\ell+3/2).
\end{equation}
The Regge slope for a heavy-light meson is half
of the one for a light-light meson. As it was shown in
Ref.~\cite{Ols94}, it is in agreement with experimental observations.
Again, the correct Regge slope in this case is not $4 a$ but $\pi a$
\cite{Ols94}.
With the same rescaling as above, $a=\pi\sigma/4$, the approximate
result~(\ref{masshl0}) can be greatly improved.

Equation~(\ref{masshl0}) can also be applied to a special type of heavy
hybrid mesons called hybrid gluelumps,
that is a pointlike heavy $q\bar q$ pair with mass $m_{q\bar q}$ bound
to a constituent gluon. We have then
\begin{equation}\label{masshl}
  (M_{hyb}-m_{q\bar q})^2=4\lambda(2n+\ell)+6\lambda,
\end{equation}
where $\lambda$ is the string tension between the gluon and the
$q\bar q$ pair. Note that $n$ and $\ell$ are the quantum numbers of the
gluon, since the dynamics of the heavy $q\bar q$ pair can be neglected.
Such a Regge-like behavior for heavy hybrid mesons has been obtained
numerically in Ref.~\cite{gluelump}.

\section{Hybrid mesons and the excited flux tube}\label{hybexci}

It is generally accepted that the static potential between the quark and
the antiquark in an usual meson is compatible with a funnel potential,
\begin{equation}\label{pot_qq}
  V_F = ar - \frac{4\alpha_S}{3r},
\end{equation}
where $\alpha_S$ is the strong coupling constant. The Coulomb part comes
from one-gluon exchange process, while the $ar$ part is a pure
confinement: it corresponds to the (classical) energy of a straight
string linking the quark and the antiquark, and whose
energy density is $a$. The description of the confining interaction in
terms of such a string is an effective approach which can
be derived from QCD \cite{dubi94}. Let us note that the spectrum
obtained with potential~(\ref{pot_qq}) is in good agreement with
experimental data for the light and heavy mesons \cite{truc}. Typical
values for the parameters fitting the lattice QCD data are
$a\approx0.2$ GeV$^2$, and $\alpha_S\approx 0.2$-0.3.

The nonabelian nature of QCD makes possible for a gluon to
interact with other gluons. These kind of self-interactions allow the
gluonic field to be in an excited state, like it is expected to be the
case in hybrid mesons. These excitations can be translated in a stringy
language by computing the energy spectrum of a quantum string, the
string fluctuations corresponding to the color field excitations. The
string energy, and consequently the potential energy between the static
quark and antiquark is then given by \cite{arvi}
\begin{equation}\label{pothyb}
  V(r)=\sqrt{a^2r^2+2\pi aN+{\cal E}^2},
\end{equation}
where $N$ is the string excitation number. Let us note that the values
of this number are constrained by the fact that the quantity under the
square root must be positive.

It is generally accepted in string theory that
${\cal E}^2=-2\pi a(D-2)/24$, with $D$ the dimension of space. Together
with $D=26$, this value indeed ensures that the Lorentz invariance is
still present at the quantum level. It is worth mentioning that we are
here dealing with effective models of QCD with $D=4$: our string is not
a fundamental object but an effective one arising from the exchanged
gluons. Consequently, it is clear that such an effective model could be
characterized by a nonstandard value of ${\cal E}$. Moreover, a
potential model like the SSH is \textit{a priori} noncovariant since
the
potentials which are used are instantaneous: the Lorentz invariance is
already broken at the classical level, so one does not need to restrict
${\cal E}$ to the usual value. From an effective model point of view,
the best interpretation to give to ${\cal E}$ seems thus the one
proposed in Ref.~\cite{brink}, where it is shown that it represents the
zero point energy of the transverse string fluctuations. In a QCD model,
this could be the zero point energy of the gluonic field. We will turn
to this interpretation later, to see that it is indeed compatible with
our results. Let us note that for large $r$, potential~(\ref{pothyb}) is
approximately equal to
\begin{equation}\label{luterm}
  V(r)=ar+\frac{\pi}{r}\left(N_L+\frac{{\cal E}^2}{2\pi a}\right).
\end{equation}
In $D=4$, the usual string theory states that ${\cal E}^2/2\pi
a=-1/12$: we recover in this limit the well-known universal L\"uscher
term \cite{luscher}. In formula~(\ref{luterm}), the minimal allowed
value for $N_L$ is zero. In this case, $V(r)$ is expected to reproduce
the funnel potential and thus represents the interaction of a quark and
an antiquark in an ordinary meson.

We can apply the auxiliary field formalism to compute a hybrid
meson spectrum in the same way as it was done in the previous section
for mesons. However, the potential to use is now given by
formula~(\ref{pothyb}) instead
of the usual $ar$. Then, the hybrid masses are given by
\begin{equation}
  M_{hyb}(\mu_i,\nu)=M(\mu_i,\nu)+\frac{2\pi a N+{\cal E}^2}{2\nu},
\end{equation}
with $M(\mu_i,\nu)$ given by Eq.~(\ref{mass2af}). Because of the new
term, $\nu$ is more complex to eliminate. Nevertheless, we can readily
compute that, in the case of a heavy hybrid meson with the quark and the
antiquark of the same mass, we have $M(\mu_i,\nu)\approx2m+\nu/2$ (the
dynamical contribution of the quark and of the antiquark is neglected),
and consequently,
\begin{equation}\label{regge_hyb}
  (M_{hyb}-2m)^2\approx 2\pi a N+{\cal E}^2.
\end{equation}
This is a kind of Regge trajectory with respect to the string excitation
number $N$. It is interesting to notice that this formula is analog to
the result of Eq.~(\ref{masshl}) concerning the heavy hybrid mesons,
with
$m_{q\bar q}=2m$, $\lambda=2 a$ and the proper rescaling $a=\pi\sigma/4$
(these values for $m_{q\bar q}$ and $\lambda$ will be justified in
Sec.~\ref{hhh}). The quantum numbers of the gluon, $(2n+\ell)$, are
replaced by $N$ the string excitation number, and the square zero point
energy of the system, $6\lambda$, is now replaced by ${\cal E}^2$, the
square zero point energy of the string fluctuations. This correspondence
is a first hint that the description of heavy hybrid mesons in terms of
a constituent gluon or of an excited string possesses a similar physical
content, expressed in terms of two different degrees of freedom, namely
a gluon or a string excitation. We will further investigate this
correspondence in the following.

\section{Hybrid mesons with constituent gluons}\label{hybcons}

\subsection{The three-body problem}

In this picture, it is assumed that the excitations of the gluon field
can be described by a constituent gluon interacting with the
quark-antiquark pair. This pair is thus in a color octet in order for
the hybrid meson to be a colorless object. Assuming the Casimir scaling
hypothesis, which seems to be confirmed by several models
\cite{scaling}, it can be shown that the confinement is no more a
Y-junction like in a baryon but two fundamental strings linking each
quarks to the gluon \cite{gluelump}. Neglecting all the short-range
interactions, the three-body SSH is thus
\begin{equation}\label{mainH}
  H=\sum_{i=q,\bar q,g}\sqrt{\bm p^2_i+m^2_i}+\sum_{j=q,\bar q}a|\bm
  x_j-\bm x_g|,
\end{equation}
with $m_g=0$. Introducing the AF, we obtain, in analogy with
Hamiltonian~(\ref{ham2}),
\begin{eqnarray}\label{ham3b1}
  H(\mu_i,\nu_i)&=&\sum_{i=q,\bar q,g}\left[\frac{\bm p^2_i+m^2_i}{2
  \mu_i}
  +\frac{\mu_i}{2}\right]\nonumber \\
  &&+\sum_{j=q,\bar q}\left[\frac{a^2(\bm x_j-\bm
  x_g)^2}{2\nu_j}+\frac{\nu_j}{2}\right].
\end{eqnarray}
We can then apply the procedure described in Ref.~\cite{coqm}, where we
solved the three-body covariant oscillator quark model by an appropriate
change of variables.

First of all, we will replace the
quark coordinates
$\bm{x}_{i}=\left\{\bm{x}_{g},\bm{x}_{q},\bm{x}_{\bar q}\right\}$ by
$\bm{x}'_{k}=\left\{\bm{R},\bm{r},\bm{y}\right\}$, with the center
of mass defined as
\begin{equation}\label{cmdef}
\bm{R}=\frac{\mu_{g}\bm{x}_{g}+\mu_{q}\bm{x}_{q}+\mu_{\bar q}\bm{x}_{
\bar q}}{\mu_{t}}.
\end{equation}
$\mu_{t}=\mu_{g}+\mu_{q}+\mu_{\bar q}$ and $\{\bm{r},\bm{y}\}$ are two
relative
coordinates. The change of coordinates is made via a matrix $Q$, thanks
to the relation $\bm{x}_{i}=Q_{ik}\bm{x}'_{k}$. Let us note that the
invariance of the Poisson brackets demands that
$\bm{p}_{i}=\left(Q^{-1}\right)^{\rm{T}}_{ik}\bm{p}'_{k}$, with
$\bm{p}'_{i}=\left\{\bm{P},\bm{p},\bm{p}_{y}\right\}$. We
define
\begin{equation}
Q=\left( \begin{array}{ccc}
 1 & A & B\\
 1 & C & D\\
 1 & E & F
 \end{array}\right),
\end{equation}
and choose to impose the constraints
\begin{equation} \label{c1}
  D=F,\quad C=E+1,
\end{equation}
in order to have a clear physical meaning for $\bm r$, that is simply
\begin{equation}
\bm r=\bm x_q-\bm x_{\bar q}.
\end{equation}
Moreover, we ask that
\begin{subequations}\label{equt}
 \begin{eqnarray}
A&=&-\frac{\mu_{q}}{\mu_{g}}C-\frac{\mu_{\bar q}}{\mu_{g}}E,\label{eq1}
\\
B&=&-\frac{\mu_{q}}{\mu_{g}}D-\frac{\mu_{\bar q}}{\mu_{g}}F,\label{eq2}
\\
E&=&-\frac{\mu_q}{\mu_q+\mu_{\bar q}},\label{eq3}\\
F&=&\frac{\mu_g}{\sqrt{\mu_t(\mu_q+\mu_{\bar q})}}\label{eq4}.
\end{eqnarray}
\end{subequations}
Constraints~(\ref{eq1}) and (\ref{eq2}) are consequences of the
definition~(\ref{cmdef}): they allow the vanishing of the terms in
$\bm P\cdot\bm p$ and $\bm P\cdot\bm p_y$ when
Hamiltonian~(\ref{ham3b1}) is rewritten in the new coordinates.
Equation~(\ref{eq3}) ensures that the cross product $\bm p\cdot\bm p_y$
is equal to zero too. In the general case, these constraints are not
sufficient to eliminate the terms in $\bm r\cdot\bm y$. However, if
$m_q=m_{\bar q}=m$, that is to say that $\mu_q=\mu_{\bar q}=\mu$ and
$\nu_q=\nu_{\bar q}=\nu$, these terms vanish. In what follows, we will
thus restrict ourselves to the case of a quark and an antiquark with the
same mass. In the center of mass frame, $\bm P=\bm 0$, the
Hamiltonian~(\ref{ham3b1}) becomes
\begin{equation}\label{ham3b2}
  H(\mu_i,\nu)=\frac{\bm p^2}{2\tilde\mu}+\frac{\bm p^2_y}{2\mu_g}+\frac
  {1}{2}\Omega_r \bm r^2+\frac{1}{2}\Omega_g \bm y^2+\frac{m^2}{\mu}+\mu
  +\frac{\mu_g}{2}+\nu,
\end{equation}
with
\begin{equation}
  \Omega_r=\frac{a^2}{2\nu},\quad\Omega_g=\frac{a^2(\mu_g+2\mu)}
  {\mu\nu},
\end{equation}
and $\tilde\mu=\mu/2$. As in Ref.~\cite{coqm}, our transformation leads
to a Hamiltonian were all variables are separated. Actually, we have
decoupled the three-body hamiltonian~(\ref{mainH}) into a sum of two
Hamiltonians: one for the two-body quark-antiquark system, the $\bm r$
dependent part, and one for the gluon, the $\bm y$ dependent part. To
confirm this point, let us mention that the mass term appearing with
$\bm p^2$ is the reduced mass of the quark and the antiquark, and the
one appearing with $\bm p^2_y$ is $\mu_g$. This nice feature is due to
our choice of $F$, given by Eq.~(\ref{eq4}). Moreover, $A=0$ when
$\mu_q=\mu_{\bar q}$. This means that the gluon position is only given
by a function of $\bm y$. But, this separation is only formal, since the
AF still have to be eliminated. This will make appear the couplings
between the three bodies.

Before doing this, we can remark that Hamiltonian~(\ref{ham3b2}) is
the sum of two harmonic oscillators. The mass spectrum and wave
functions are then easily obtained. They read
\begin{equation}\label{en3b}
  E(\mu_i,\nu)=\omega_r(2n+\ell+3/2)+\omega_g({\cal N}+3/2)+\frac{m^2}{
  \mu}+\mu+\frac{\mu_g}{2}+\nu,
\end{equation}
\begin{eqnarray}\label{fo}
\psi&=&\psi_{q\bar q}(\bm r)\times\psi_{g}(\bm
y)=\phi_{n,\ell}(\beta^{1/2}_r\, r)Y^{
m}_{\ell}(\theta,\varphi)\nonumber\\
&&\times \phi_
{n_{y},\ell_{y}}(\beta^{1/2}_y\, y)Y^{m_{y}}_{\ell_{y}}(\theta_{y},
\varphi_{y}),
\end{eqnarray}
and
\begin{equation}
  \omega_r=\frac{a}{\sqrt{\mu\nu}},\quad\omega_g=
  a\sqrt{\frac{\mu_g+2\mu}{\mu\nu\mu_g}},
\end{equation}
\begin{equation}
  \beta_r=\tilde\mu\omega_r,\quad\beta_g=\mu_g\omega_y.
\end{equation}
For later convenience, we also defined
\begin{equation}
  {\cal N}=2n_y+\ell_y.
\end{equation}
The allowed values for ${\cal N}$ are 0, 1, \ldots It is worth noting
that the state ${\cal N}=0$ does not correspond to an ordinary meson
state (as in the case $N_L=0$ for the potential~(\ref{luterm})), but to
the hybrid meson ground state.

Formulas~(\ref{en3b}) and (\ref{fo}) are a generalization of the results
of Ref.~\cite{hyb1}, since in this last work the quark and the antiquark
were assumed to be fixed and the dynamics of the system was actually a
one-body problem. Here, we deal with the full three-body system. Using
the well-known properties of the harmonic oscillator, it is easy to
compute the quantities
\begin{equation}
\left\langle r^2\right\rangle =\frac{(2n+\ell+3/2)}{\beta_r},\quad\left
\langle y^2\right\rangle=\frac{({\cal N}+3/2)}{\beta_g}.
\end{equation}
They give information about the geometric configuration of the three
bodies. The elimination of the AF has now to be performed by minimizing
the energy~(\ref{en3b}) with respect to them. This problem leads to
rather complex expressions in general, but can be simplified in the case
of heavy and light hybrids.

\subsection{Heavy hybrid mesons}\label{hhh}

In this section, we will consider that $m\gg \sqrt{a}$ to obtain simple
analytical formula. In this case, the quark and the antiquark are very
heavy and we can set $\mu\approx m$. Formula~(\ref{en3b}) then reduces
to
\begin{eqnarray}
  E(\mu_g,\nu)&=&\frac{a}{\sqrt{m\nu}}(2n+\ell+3/2)+a\sqrt{\frac{2}{\nu
  \mu_g}}({\cal N}+3/2)\nonumber\\
  &&+2m+\frac{\mu_g}{2}+\nu.
\end{eqnarray}
Neglecting the excitation energy of the $q\bar q$ pair, we obtain
\begin{eqnarray}
  E(\mu_g,\nu)  &\approx&a\sqrt{\frac{2}{\nu\mu_g}}({\cal N}+3/2)+2m+
  \frac{\mu_g}{2}+\nu.\label{ehh}
\end{eqnarray}
As $E(\mu_g,\nu)$ is symmetric for the exchange
$\mu_g\leftrightarrow2\nu$, we can set
\begin{equation}\label{virhh}
  \mu_g\approx2\nu
\end{equation}
to simplify the energy formula~(\ref{ehh}).
The constraint $\delta_\nu E(\nu)=0$ gives
\begin{equation}\label{nuhh}
  \nu_0=\sqrt{\frac{a}{2}({\cal N}+3/2)},
\end{equation}
\begin{equation}
  (E-2m)^2=8a{\cal N}+12a.
\end{equation}
This last mass formula is clearly analog to Eq.~(\ref{masshl}). The
string tension is here $2a$ because the total string
results in the superposition of the two fundamental strings linking the
gluon to the quark and to the antiquark, in the limit of a pointlike
$q\bar q$ pair with $m_{q\bar q}=2 m$ (no interaction between the
quarks).

As it was argued in a previous study using special relativity
arguments \cite{scz}, the geometrical configuration of a heavy hybrid
meson is
most likely to be a quark and an antiquark close to each other, with a
gluon orbiting around the pair. In our model, it is easy to compute that
in the ground state ($n=\ell={\cal N}=0$),
\begin{equation}\label{struc}
  \left\langle r^2\right\rangle=\frac{3^{5/4}}{a^{3/4}\sqrt{2m}}\ <\
  \left\langle y^2\right\rangle=\frac{3}{4a}.
\end{equation}
The quark-antiquark separation is smaller than the distance between the
gluon and the center of mass, in agreement with results of Ref.~\cite{
scz}. This is
the hybrid gluelump picture that we already mentioned in
Sec.~\ref{afintro}, and which was studied in Ref.~\cite{gluelump}. In
this case, it is also showed in Ref.~\cite{hyb1} that the heavy hybrid
masses are in agreement with lattice QCD calculations.

Let us point out that the results of this section are strictly valid in
the limit $m\rightarrow\infty$, like in the case of static quarks which
are considered in lattice QCD. Indeed, even if we set $m=5$~GeV, that is
a value slightly above the $b$ quark mass, relation~(\ref{struc}) is not
true. We have in this case
$\left\langle r^2\right\rangle\approx\left\langle y^2\right\rangle$.

\subsection{Light hybrid mesons}\label{lh}

One of the nice features of our formalism is that it is well defined for
vanishing quark masses. In this case we can assume that
$\mu\approx\mu_g$, since the gluon, the quark and the antiquark are
massless particles. Mass formula~(\ref{en3b}) then becomes
\begin{equation}
\label{enll}
  E(\mu,\nu)=\frac{a}{\sqrt{\mu\nu}}(2n+\ell+3/2)+a\sqrt{\frac{3}
  {\mu\nu}}({\cal N}+3/2)+\frac{3}{2}\mu+\nu.
\end{equation}
The symmetry of Eq.~(\ref{enll}) allows us to use the relation
\begin{equation}\label{virll}
  3\mu=2\nu.
\end{equation}
A numerical solution of mass formula~(\ref{en3b}) for $m=0$ gives
$\mu/\mu_g=0.81$ and $3\mu/2\nu=0.93$ for the ground state, whatever the
value of $a$, in
quite good agreement with our approximations. Let us note that they are
mostly valid when the excitation energies of the quarks and the gluon
are similar.

After the minimization of $E$ with respect to $\nu$ we obtain
\begin{equation}
  \nu_0=\sqrt{\frac{a}{2}\sqrt{\frac{3}{2}}(2n+\ell+3/2)+\frac{3a}{2
  \sqrt{2}}({\cal N}+3/2)},
\end{equation}
\begin{equation}\label{enll2}
  E^2=4\sqrt{6}\left[a(2n+\ell+3/2)+a\sqrt{3}({\cal N}+3/2)\right].
\end{equation}
This formula predicts that the light hybrid mesons should exhibit Regge
trajectories at large $\ell$ like the usual light mesons. The
trajectories corresponding to different values of ${\cal N}$ are
parallel
but differ in their intercept: two successive trajectories are separated
by $2^{5/2}3a$.

Concerning the structure of these hybrid mesons, we can compute
that in the ground state
\begin{equation}\label{rmll}
  \left\langle r^2\right\rangle=\frac{3}{2a}\sqrt{6}\ >\ \left\langle y^
  2\right\rangle=\frac{3}{2a}\frac{1}{\sqrt{2}}.
\end{equation}
The larger quantity is now clearly the quark-antiquark separation. The
light hybrid meson structure is thus rather different from the heavy
hybrid meson one.
In particular, the picture of a pointlike $q\bar q$ pair is no longer
valid.

We can use formula~(\ref{enll2}) and (\ref{rmll}) to estimate the
mass of the lightest hybrid meson in our model. The ground state mass is
given by $E_0=6.337\sqrt{a}$\, GeV. In a first approximation, the total
energy should be given by
$E=E_0+\Delta E$ with
\begin{equation}\label{oge1}
  \Delta E=\left\langle \frac{\alpha_S}{6r}-\frac{3\alpha_S}{2r_{qg}}-
  \frac{3\alpha_S}{2r_{\bar qg}}\right\rangle
\end{equation}
encoding the one gluon exchange processes at the lowest
order \cite{horn}. Thanks to relations~(\ref{equt}), we have
\begin{equation}\label{formu}
  \bm x_g-\bm x_q=-\sqrt{\frac{\mu_t}{\mu_q+\mu_{\bar q}}}\bm y-\frac{
  \bm r}{2}.
\end{equation}
Assuming that $|\bm x_g-\bm x_q|\approx|\bm x_g-\bm x_{\bar q}|$
thanks to the symmetry of our problem, Eq.~(\ref{rmll}) implies
\begin{equation}
  \left\langle r^2_{qg}\right\rangle=\left\langle r^2_{\bar qg}\right
  \rangle\approx\frac{\left\langle r^2\right\rangle}{4}+\frac{3\left
  \langle y^2\right\rangle}{2}=\frac{2.510}{a}.
\end{equation}
Thus,
\begin{equation}
  \Delta E\approx\frac{\alpha_S}{6\sqrt{\left\langle r^2\right\rangle}}-
  \frac{3\alpha_S}{\sqrt{\left\langle r_{qg}^2\right\rangle}}.
\end{equation}
With the value $\alpha_S=0.4$, which was already successfully used in
the description of light mesons \cite{instanton,fulch}, we find
$\Delta E=-0.724\sqrt{a}$ and $E=5.613\sqrt{a}$. We already pointed out
in Sec.~\ref{afintro} that the AF method gives qualitative results in
agreement with observations, but overestimates the masses. The Regge
slope is here $4\sqrt{6}a$. But we can expect that, at large $\ell$,
the contribution of the constituent gluon can be neglected with respect
to the contribution of the quark-antiquark pair. Then, the exact slope
should be given by $2\pi a$ as in the meson case. As it is proposed in
Sec.~\ref{afintro}, we
can thus rescale the string tension and define
$a=2\pi\sigma/4\sqrt{6}$, with $\sigma=0.2$ GeV$^2$ the physical string
tension. We finally obtain
\begin{equation}\label{elh}
E=4.495\sqrt{\sigma}=2.010\ {\rm GeV}
\end{equation}
for the lightest hybrid meson.

It is worth mentioning that several approaches, such as QCD in Coulomb
gauge \cite{llanes}, flux tube model \cite{merlin}, and lattice QCD
\cite{Bern97}, lead to the conclusion that the lightest hybrid meson
mass is around $2$ GeV. Interestingly, this is close to the recently
observed $\pi_1(2000)$ exotic state \cite{Lu}. However, we stress that,
since our model neglects the spin interactions, the energy~(\ref{elh})
is only a rough estimation of the lightest hybrid mass. As formula~(\ref
{enll2}) does not involve the spin quantum numbers, its ground state (
$n=\ell={\cal N}=0$) could have the following quantum numbers
\cite{kala}
\begin{equation}\label{multi}
  J^{PC}=0^{\pm+},1^{\pm+},1^{\pm-},2^{\pm+}.
\end{equation}
These eight states are clearly degenerate in our approach. Our
estimation of the ground state mass should thus be regarded as a spin-
averaged mass of the multiplet~(\ref{multi}). The spin corrections are
expected to contribute for at most $10\%$ of the total mass \cite{kala}.
Consequently, they have to be taken into account if one wants to make an
accurate comparison with either lattice or experimental data.

It is interesting to mention that, in the flux tube model, the lowest
hybrid states can have the following quantum numbers \cite{merlin}:
\begin{equation}\label{multif}
  J^{PC}=0^{\pm\mp},\, 1^{\pm\pm},\, 1^{\pm\mp},\, 2^{\pm\mp}.
\end{equation}
The multiplets predicted by the flux tube model and the constituent
gluon approach are not identical: only six states on eight are in
correspondence. The inclusion of spin interactions in both approaches
would thus lead to different results, but a detailed comparison of these
differences is out of the scope of this paper.

\section{Effective two-body potentials}\label{eff2pot}

After having studied the hybrid mesons as a three-body system, we would
like to connect this model with a two-body description. More precisely,
we would like to absorb the gluonic degree of freedom into an effective
potential between the quark and the antiquark. To do this, we begin by
averaging the three-body Hamiltonian~(\ref{ham3b2}) on the gluon wave
function $\left| \psi_g\right\rangle$. We assume that
\begin{eqnarray}\label{effham}
  H_{q\bar q}(\mu_i,\nu)&=&\left\langle \psi_g\right|H(\mu_i,\nu)\left|
  \psi_g\right\rangle \nonumber\\
  &=&\frac{\bm p^2+m^2}{\mu}+\mu+\frac{1}{2}\Omega_r r^2+\omega_g({\cal
  N}+3/2)\nonumber \\
  &&+\frac{\mu_g}{2}+\nu.
\end{eqnarray}
The first two terms are the kinetic part of a two-body SSH.
Consequently, the other terms are interpreted as the effective potential
between the quark and the antiquark, i. e.
\begin{equation}\label{effpot1}
  V_{q\bar q}(\mu_i,\nu)=\frac{1}{2}\Omega_r r^2+\omega_g({\cal N}+3/2)+
  \frac{\mu_g}{2}+\nu.
\end{equation}
 This potential being still dependent of the AF, we have to carefully
 remove them. This will be done in the two special cases we treated
 previously, namely the light and heavy hybrid mesons.

\subsection{Heavy hybrid mesons}\label{effh}

When $\mu=m$, the potential~(\ref{effpot1}) only depends on $\nu$ and
$\mu_g$. But, following relation~(\ref{virhh}), $\mu_g=2\nu$. This
allows to eliminate the gluonic degree of freedom and replace it by a
stringy equivalent. Doing this, $\nu$ becomes the AF associated with an
``effective" string linking the quark and the antiquark.  We have
\begin{equation}
  H_{q\bar q}(\nu)=2m+\frac{\bm p^2}{m}+V_{q\bar q}(\nu),
\end{equation}
with
\begin{equation}
  V_{q\bar q}(\nu)=\frac{a^2r^2+4a({\cal N}+3/2)}{4\nu}+2\nu.
\end{equation}
The condition $\delta_\nu H_{q\bar q}(\nu)=0$ leads after replacement to
\begin{equation}\label{effpot2}
  V_{q\bar q}=\sqrt{2}\sqrt{a^2r^2+4a({\cal N}+3/2)}.
\end{equation}
The $\sqrt{2}$ factor is clearly unphysical, since the asymptotic form
of $V_{q\bar q}$ should be $ar$ \cite{Allen:1998wp}. Actually, it is due
to the AF formalism itself, which overestimates the masses, and thus the
potential energies too. As we argued in Sec.~\ref{afintro}, a rescaling
of the string tension can be performed to find the correct expression.
If we define $a=\sigma/\sqrt{2}$, we have
\begin{equation}\label{effpot3}
  V_{q\bar q}=\sqrt{\sigma^2r^2+2^{5/2}\sigma{\cal N}+{\cal E}^2_h},
\end{equation}
with
\begin{equation}
  {\cal E}_h=2^{3/4}\sqrt{3\sigma}.
\end{equation}
A remarkable feature has to be pointed out: The effective
potential~(\ref{effpot3}) is formally equivalent to the one of the
excited flux
tube picture~(\ref{pothyb}). This draws
a strong analogy between the
gluonic and the string fluctuation degrees of freedom. The quantum
numbers of the gluon, namely ${\cal N}=2n_y+\ell_y$, are analog to the
string excitation number $N$. But the number
$\cal N$ can always take the value 0 since ${\cal E}_h$ is positive.
Moreover, as ${\cal E}$ was interpreted as
the zero point energy of the transverse string fluctuations, we can
interpret ${\cal E}_h$ as the zero point energy of the string-gluon
system. Indeed, following relations~(\ref{virhh}) and (\ref{nuhh}), this
zero point energy, associated with the gluon and the two fundamental
strings, is given by
\begin{equation}
\left.  2\nu+\mu_g\right|_{{\cal N}=0}=2\sqrt{3a}=2^{3/4}\sqrt{3\sigma}=
{\cal E}_h.
\end{equation}

This shows that a constituent gluon linked to a heavy quark and a heavy
antiquark by two fundamental strings is equivalent to an excited string
linking the quark and the antiquark. This string is an effective one;
its quantum numbers and zero energy are those of the corresponding
gluonic field. Let us also remark that $2^{5/2}=5.66$, which is around
the $2\pi$ factor obtained with string theory (see
formula~(\ref{pothyb})).

An estimation of the constituent gluon mass is here given by
$\mu_g={\cal E}_h/2$. For the standard value $\sigma=0.2$~GeV$^2$, we
obtain $0.651$~GeV. This value is close to the usual ones used in
potential models \cite{coqm,gluelump}.

\subsection{Light hybrid mesons}\label{effl}

In the case of light hybrids, the situation is slightly more complex.
Equation~(\ref{virll}) clearly suggest to replace $\mu_g$ by $2\nu/3$.
This can be done in analogy with the heavy hybrid case. However, in the
present case, the Hamiltonian is
\begin{equation}
  H_{q\bar q}(\mu,\nu)=\mu+\frac{\bm p^2}{\mu}+V_{q\bar q}(\mu,\nu).
\end{equation}
Consequently, $\mu$ remains present in the effective potential.
Strictly speaking, the effective potential always depends on the
$q\bar q$ state, but this dependence drops for heavy quarks since
$\mu\approx m$. $V_{q\bar q}(\mu,\nu)$ reads
\begin{equation}
  V_{q\bar q}(\mu,\nu)=\frac{a^2r^2}{4\nu}+\frac{a}{\nu}\sqrt{\frac{\nu+
  3\mu}{\mu}}({\cal N}+3/2)+\frac{4\nu}{3}.
\end{equation}
The elimination of $\mu$ and $\nu$ cannot then be performed
analytically. What can be done however is to find the asymptotic
expression of the effective potential. As $V_{q\bar q}$ has to grow like
$ar$ for large $r$, we have indeed
\begin{equation}
  V_{q\bar q}(\mu,\nu)\approx
  V^0(\nu)=\frac{a^2r^2}{4\nu}+\frac{4\nu}{3}.
\end{equation}
The minimization of $V^0$ with respect to $\nu$ gives
\begin{equation}
  \nu_0=\frac{\sqrt{3}}{4}ar,\quad  V^0=\frac{2}{\sqrt{3}}\, ar.
\end{equation}
The scaling $a=\sqrt{3}\sigma/2$ provides the correct behavior in
$\sigma r$.

The first correction to this potential is given by
\begin{equation}
  \Delta V=\frac{a}{\nu}\sqrt{\frac{\nu+3\mu}{\mu}}({\cal N}+3/2)=\frac{
  2}{r}\sqrt{\frac{\sigma r+8\mu}{2\mu}}({\cal N}+3/2).
\end{equation}
This term is a kind of generalization of the L\"uscher
term~(\ref{luterm}). The elimination of $\mu$ from the Hamiltonian
\begin{equation}
  H^0(\mu)=\frac{\bm p^2}{\mu}+\mu+V^0
\end{equation}
gives \cite{qse}
\begin{equation}
  \mu_{n\ell}=\sqrt{\sigma}\left(\frac{\epsilon_{n\ell}}{3}\right)
  ^{3/4},
\end{equation}
where $\epsilon_{n\ell}$ is an eigenvalue of the dimensionless operator
$(\bm q^2+|\bm x|)$. Approximated analytical formula for
$\epsilon_{n\ell}$ can be found in Refs.~\cite{qse,bose}. As $\mu$
increases with $n$ and $\ell$, $\Delta V$ becomes in this
limit very similar to the L\"uscher term \cite{luscher}:
$\Delta V\approx4  ({\cal N}+3/2)/r$. We see that the light hybrid
mesons are complex systems, and that the corresponding effective
two-body potential is not so easily obtained than in the heavy hybrid
meson case. Such a study deserves numerical computations that we leave
for future works.

\section{Comparison with lattice QCD}\label{discuss}

One of the observables in lattice QCD is the potential energy between a
static quark-antiquark pair. Such a pair can be identified with the
$q\bar q$ pair in
heavy hybrid mesons. It appears that there are several levels of
potential energy, corresponding to different states of the gluon field
\cite{Juge}. These excited states of the gluonic field are labeled by
three quantum numbers. The first one is the excitation number ${\rm N}$.
The second one is the magnitude of the projection $\Lambda$ of the total
gluon field momentum $\vec J_g = \vec L_g + \vec S_g$ on the $q\bar q$
axis. The capital Greek letters $\Sigma,\Pi,\Delta,\ldots$ are used to
indicate the states with $|\Lambda |= 0,1,2,\ldots$ respectively. The
combined operations of the parity and the C-parity
is also a symmetry. Its eigenvalue is
denoted by $\eta_{CP}$. States with $\eta_{CP}=1 (-1)$ are denoted by
the subscripts $g$ ($u$). There is a additional label for the $\Sigma$
states: $\Sigma$ states which are even (odd) under a reflection in a
plane containing the $q\bar q$ axis are denoted by a superscript $+$
($-$). Many different states have been computed in Ref.~\cite{Juge2}.
Let us note that the excitation number ${\rm N}$ used in
Ref.~\cite{Juge2} is linked to ours by ${\rm N}={\cal N}+1$ and to the
L\"uscher number by ${\rm N}=N_L$. Let us
remind that in our model, the number ${\cal N}=0$ corresponds to a
hybrid meson ground state, while the number ${\rm N}=0$ corresponds to
an ordinary meson.

\begin{figure}
  \includegraphics[width=8cm]{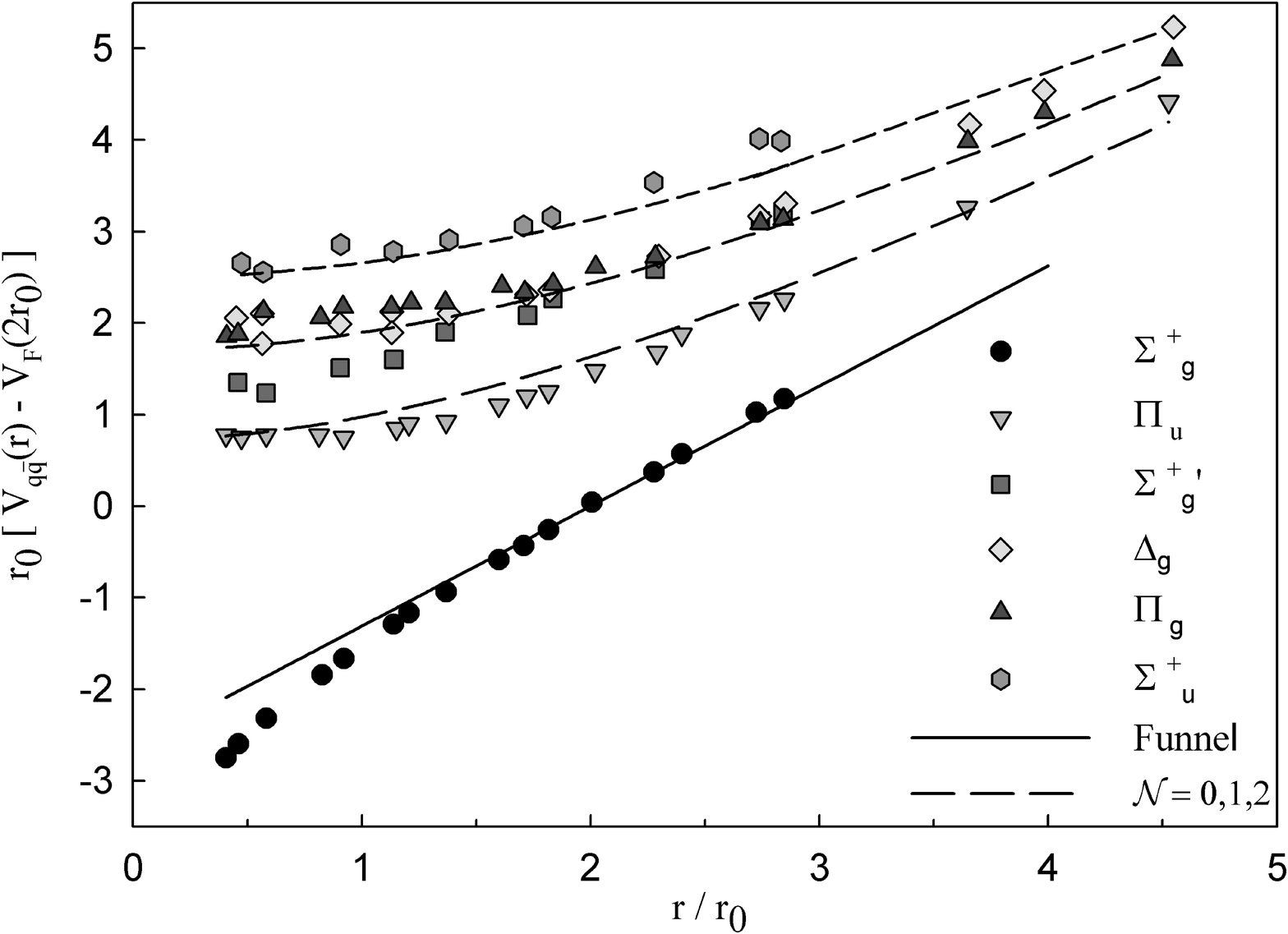}
\caption{Comparison between lattice QCD calculations (symbols) from
Ref.~\cite{Juge}, and our effective potential
$V_{q\bar q}$~(\ref{effpot4})
with different values of ${\cal N}$ (dotted lines) for a heavy hybrid
meson. The parameter
$\sigma=0.21$ GeV$^{2}$ is fitted on ground state $\Sigma_g^+$, with the
Funnel potential $V_F=\sigma r-4\alpha_S/3r$. All the potentials are
plotted in terms of the lattice scale $r_0 = 2.5$ GeV$^{-1}$ and are
shifted by an overall amount $V_F(2r_0)$. Only the confining part of the
funnel potential is plotted in this figure. To clarify the graph, only
one of the four lattice states with ${\cal N}=2$ was plotted.}
\label{fig:hyb1}
\end{figure}

\begin{figure}
 \includegraphics[width=8cm]{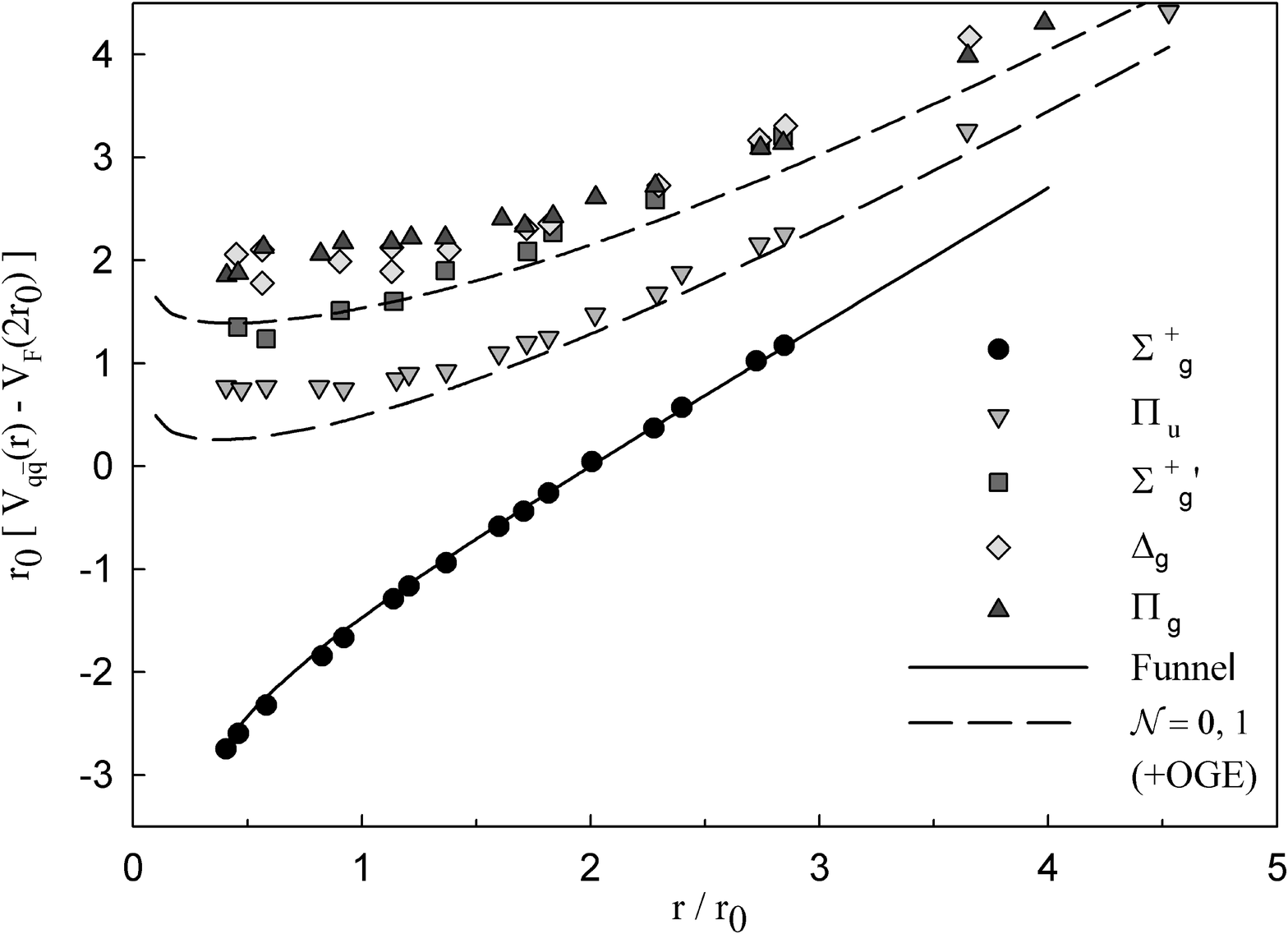}
\caption{Same as Fig.~\ref{fig:hyb1} but the short-range part
$\Delta V$~(\ref{oge3}) is added to $V_{q\bar q}$~(\ref{effpot4}).
$\alpha_S=0.234$
is obtained from a fit of the funnel potential on the ground state.}
\label{fig:hyb2}
\end{figure}

We can compare the energy levels of lattice QCD with those
predicted by our model in the limit of heavy hybrid mesons (infinite
quark mass). Following Eq.~(\ref{effpot3}), the effective two-body
potential in a hybrid meson is
\begin{equation}\label{effpot4}
  V_{q\bar q}=\sqrt{\sigma^2r^2+2^{5/2}\sigma({\cal N}+3/2)}.
\end{equation}
We can see in Fig.~\ref{fig:hyb1} that the simple
expression~(\ref{effpot4}) fits rather well the lattice data. However,
this potential is
a pure confinement. The simplest way to include a short range
interaction is to add to $V_{q\bar q}$ the effective one-gluon exchange
potential
\begin{equation}
  \Delta V=\left\langle \psi_g\right| \frac{\alpha_S}{6r}-\frac{3
  \alpha_S}{2r_{qg}}-\frac{3\alpha_S}{2r_{\bar qg}}\left|\psi_g\right
  \rangle.
\end{equation}
In analogy with formula~(\ref{oge1}) and using Eq.~(\ref{formu}), we can
approximately compute $\Delta V$, and we obtain
\begin{equation}\label{oge3}
  \Delta
  V\approx\frac{\alpha_S}{6r}-\frac{3\alpha_S}{\sqrt{\frac{r^2}{4}+\frac
  {({\cal N}+3/2)}{\sqrt{2}\sigma}}}.
\end{equation}
We expressed our results in terms of $\sigma$ instead of $a$ since
$\sigma$ is the physical string tension, and it has to be used instead
of $a$ in the wave function too. As illustrated in Fig.~\ref{fig:hyb2},
the addition of the contribution~(\ref{oge3}) lowers the potential
energy, but the first
states are still correctly described. This last result is in agreement
with Ref.~\cite{hyb1}.

As our model only depends on ${\cal N}$, it is not able to
reproduce the splittings between various potentials at short distances
\cite{Juge2}. However, the
constituent gluon picture can provide an intuitive explanation for this
fine structure. Let us consider the ${\cal N}=0$ state. As
${\cal N}=2n_y+\ell_y$, the only possibility is $n_y=\ell_y=0$, which
corresponds to the $\Pi_u$ state. For ${\cal N}=1$, we can only have
$n_y=0$ and $\ell_y=1$. A possible mechanism to explain the short
distance splitting of this state into three levels could be found in the
relativistic corrections to the coulomb or the confining terms, that we
neglected here. In
particular, one of these corrections is a spin-orbit term proportional
to $\bm L_g\cdot \bm S_g$ \cite{soon}. Consequently, for a nonvanishing
value of $\bm L_g$ (thus of $\ell_y$), and since the gluon is a spin $1$
particle, this spin-orbit term will split a level into three levels
defined by their value of $\bm J_g=\bm L_g+\bm S_g$. Finally, for
${\cal N}=2$, a detailed lattice study reveals four levels \cite{Juge2}.
We expect that one of them corresponds to $n_y=1$, $\ell_y=0$, and that
the three others are $n_y=0$, $\ell_y=1$ states with the spin-orbit
interaction separating them. To check this point, we need to include the
spin structure of the gluon into the computation of the effective
potential. This cannot be done analytically, and it is leaved for future
works.

\section{Conclusions}\label{conclu}

In this work, we studied two different pictures of hybrid mesons, which
are both based on a spinless Salpeter Hamiltonian with a linear
confinement. In particular, we applied the auxiliary fields technique to
obtain analytical mass formula and wave functions of our models.

The first framework describing a hybrid meson is the excited flux tube.
It relies on the idea that the flux tube (a Nambu-Goto string) linking
the quark and the antiquark is not in its ground state, but in an
excited one due to possible quantum fluctuations of the string. The
excited string approach has been widely discussed in the literature (see
for example Refs.~\cite{Allen:1998wp,luscher}). In particular, it can be
shown that the interquark confining potential in this approach is of the
form $V(r)=\sqrt{a^2r^2+2\pi aN+{\cal E}^2}$. Fundamental string theory
states that ${\cal E}^2=-\pi a/6$ to preserve the Lorentz invariance at
the quantum level. However, we argued in this study that since we are
dealing with an effective string theory, we should rather look at the
physical content of ${\cal E}^2$. As it can be interpreted as the square
zero point energy of the string fluctuations, we proposed to consider
${\cal E}$ as the square zero point energy of the gluonic field, which
is simulated in a simplified way by the string. Consequently, we are led
to the conclusion that the usual value of ${\cal E}^2$ is not the best
one for our purpose.

A second picture assimilates the hybrid meson to a three-body
quark-antiquark-gluon bound state. The constituent gluon is then linked
to the quark and the antiquark by two fundamental strings \cite{
horn,hyb1}. Thanks to the auxiliary fields technique, we have been able
to find an analytic expression for this three-body system in the case of
heavy and light hybrid mesons. In this last case, we found for the mass
scale
of the lightest hybrid mesons a value close to $2$~GeV. This is in
agreement with
other effective models \cite{llanes,merlin} and with lattice QCD
computations \cite{Bern97}.

An interesting question is: how could the constituent gluon approach be
reduced to a two-body model (only the quark and the antiquark) with an
effective potential simulating the effect of the constituent gluon? In
the heavy hybrid meson sector, we showed that the effective potential
has the form of the excited flux tube interaction, with the gluon
quantum numbers $(2 n_g+\ell_g)$ corresponding to the string excitation
number $\mathcal{N}$. Moreover, the zero point energy of the excited
flux tube,
denoted as ${\cal E}$, is equal to the zero point energy of the heavy
$q\bar q g$ system, when the $q\bar q$ energy is subtracted.
Consequently, the constituent gluon picture is in this case equivalent
to an effective string theory. In the light hybrid meson sector, the
effective potential crucially depends on the quark-antiquark state, and
only an asymptotic expression can be derived. This asymptotic expression
is similar to the L\"uscher term, but is now state dependent. It becomes
universal only for highly excited quark-antiquark states.

Finally, we compared our results with lattice QCD predictions concerning
the gluonic field energy levels \cite{Juge, Juge2}. We find a good
general agreement, but our model fails to describe the fine structure
appearing at short distances. We argued that it was due to the fact that
we did not take into account the gluon spin. Indeed, we showed by
intuitive arguments that the spin-orbit interaction of the gluon should
be able to roughly explain this fine structure, at least for the first
excited states. The inclusion of the spin structure of the $q\bar qg$
system is thus a very interesting problem, which requires further
investigations. Such a work is in progress.

\acknowledgments
The authors thank the FNRS for financial support.

\end{document}